\documentclass[10 pt, conference]{IEEEtran}  

\pdfoutput=1

\usepackage{graphicx} 
\usepackage{epsfig} 
\usepackage{amsmath} 
\usepackage{amssymb}  
\usepackage{subcaption}

\title{\LARGE \bf
Deep learning control of artificial avatars in group coordination tasks
}

\author{Maria Lombardi$^{1}$, Davide Liuzza$^{2}$ and Mario di Bernardo$^{3}$
\thanks{$^{1}$ M. Lombardi is with the Department of Engineering Mathematics,
        University of Bristol, UK
        {\tt\small maria.lombardi@bristol.ac.uk} 
        and with the Department of Electrical Engineering and Information Technology,
    	University of Naples ``Federico II", Italy
		{\tt\small maria.lombardi@unina.it}}%
\thanks{$^{2}$ D. Liuzza is with the Department of Engineering,
        University of Sannio, Benevento, Italy
        {\tt\small davide.liuzza@unisannio.it}}%
\thanks{$^{3}$ M. di Bernardo is with the Department of Engineering Mathematics,
	University of Bristol, UK
	{\tt\small m.dibernardo@bristol.ac.uk} 
	and with the Department of Electrical Engineering and Information Technology,
	University of Naples ``Federico II", Italy
	{\tt\small mario.dibernardo@unina.it}}%
}

\begin{document}

\maketitle
\thispagestyle{empty}
\pagestyle{empty}

\begin{abstract}
In many joint-action scenarios, humans and robots have to coordinate their movements to accomplish a given shared task. Lifting an object together, sawing a wood log, transferring objects from a point to another are all examples where motor coordination between humans and machines is a crucial requirement.
While the dyadic coordination between a human and a robot has been studied in previous investigations, the multi-agent scenario in which a robot has to be integrated into a human group still remains a less explored field of research. 
In this paper we discuss how to synthesise an artificial agent able to coordinate its motion in human ensembles. Driven by a control architecture based on deep reinforcement learning, such an artificial agent will be able to autonomously move itself in order to synchronise its motion with that of the group while exhibiting human-like kinematic features. As a paradigmatic coordination task we take a group version of the so-called mirror-game which is highlighted as a good benchmark in the human movement literature.
\end{abstract}

\section{INTRODUCTION}
\label{sec:introduction}
Interest in using robots and artificial avatars in joint tasks with humans to reach a common goal is growing rapidly. Indeed, it is possible to find numerous applications that span from industrial tasks to entertainment, from navigation to orientation and so on, in which artificial agents are required to interact cooperatively with people. Examples of human-robot interaction include the problem of jointly handling an object \cite{Edsinger2007}, sawing a log of wood \cite{Peternel2014}, managing a common work-piece in a production system \cite{Faber2015}, or performing a ``pick and place'' coordination task \cite{Lamb2017}. While dyadic coordination between humans and robots is the subject of much ongoing research, the problem of having robots or avatars interacting with a human team remains a seldom investigated field. This is probably due to the complex mechanisms underlying interpersonal coordination, the different ways in which coordination can emerge in human groups, and the potentially large amount of data to be collected and processed in real-time.

From a control viewpoint, the emergence of multi-agent synchronisation while performing a joint task is a phenomenon characterised by \textit{non-linear dynamics} in which an individual has to \textit{predict} what others are going to do and \textit{adjust} his/her movements in order to complement the movements of the others in order to achieve precise and accurate \textit{temporal correspondence} \cite{Vesper2010}.

In this context, an open question is to investigate whether it is possible to influence the emergent group behaviour via the introduction of artificial agents able be accepted in a natural way by the group and help it achieving a collective control goal. To reach such a goal, the artificial agent has to be able to integrate its motion with that of the others exhibiting at the same time typical human-like kinematic features. In this way, the artificial agent can merge with the rest of the group and enhance, rather disrupt, social attachment between its members and group cohesiveness.

The problem is crucial in social robotics, where new advancements in human-robot interaction can promote novel diagnostic and rehabilitation strategies for patient suffering from social and motor disorders \cite{Amado2016}.

In this work we define a ``human-like" motion by using the concept of ``Individual Motor Signature" (IMS), proposed in \cite{Slowinski2016} as a valid biomarker able to capture the peculiarity of the human motion. Specifically, the IMS has been defined in terms of the probability density function (PDF) of the velocity profiles characterising a specific joint task. 

The aim of this paper is to present a control architecture based on deep learning to drive an artificial agent able in performing a joint task in a multi-agent scenario while exhibiting a desired IMS. As a scenario of interest, we take a multiplayer version of the mirror game proposed in \cite{Noy2011} as a paradigmatic task of interpersonal motor coordination. In our version of the mirror game, first proposed in \cite{Slowinski2016}, a group of players is asked to oscillate a finger sideways performing some interesting motion and synchronising theirs with that of the others (see \cite{Alderisio2017} for further details).



The approach we follow is an extension to groups of the strategy we presented in \cite{Zhai2017,Lombardi2018a,Lombardi2018b} in the case of dyadic interactions. Specifically, in \cite{Lombardi2018b} we designed an autonomous cyber player able to play dyadic leader-follower sessions of the mirror game with different human players. Such a cyber player was driven by a  Q-learning algorithm aiming at exhibiting the kinematic features of a target human player in order to emulate hers/his way of moving when engaged in a dyadic interaction. 


Extending the Q-learning approach to multi-agent systems is cumbersome as the approach is unscalable with the growth of the system state space due to the addition of other players. To overcome this limitation, we use ``deep reinforcement learning" \cite{Sutton2018,DeepMind2015,Li2018}, combining the reinforcement learning strategy with the powerful generalization capabilities of neural networks.
To design the control architecture of the cyber player, we collect motor measurements signals of four different players involved in a joint oscillatory task and then train the CP to mimic the way of moving of one of them. The validation is done replacing the target player with the cyber player and comparing the group performance in order to prove the effectiveness of the proposed control approach.

\section{PRELIMINARIES}
\label{sec:background}

A group of people interacting with each other can be described as a complex network system in which each individual is represented as a node (or agent) with its own dynamics while the visual coupling with the other members of the group as edges in a graph describing the network of their interactions. The structure of the interconnections established among the groups' members is formalised by the adjacency matrix $A := a_{ij}$, in which the element $a_{ij} = 1$ only if the node $j$ is linked with the node $i$, or in other terms if the individual $j$ is visually coupled with the individual $i$. 

\begin{figure}[thpb]
    \centering
    \includegraphics[width=\columnwidth]{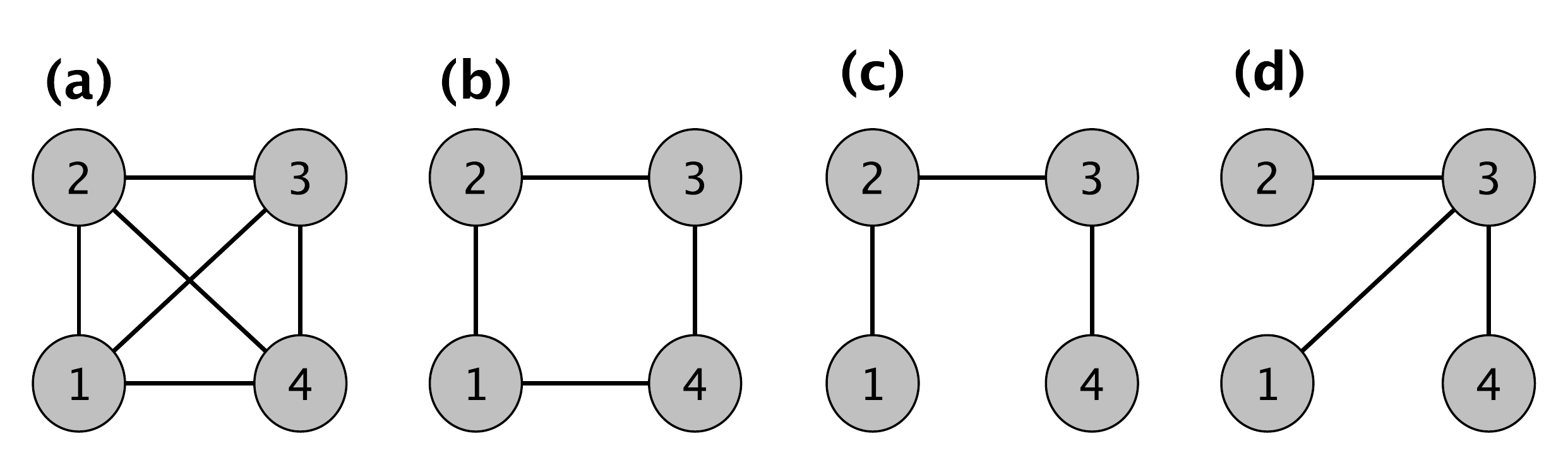}
    \caption{Network topologies. (\textbf{a}) Complete graph. (\textbf{b}) Ring graph. (\textbf{c}) Path graph. (\textbf{d}) Star graph.}
    \label{fig:topologies}
\end{figure}

Four different topologies are considered in this work as shown in Fig. \ref{fig:topologies}.
As described in \cite{Alderisio2017} these different topologies can be implemented experimentally by changing the way in which participants sit with respect to each other and by asking them to wear appropriate goggles restricting their field-of-view.

Let $\theta_k\left(t\right)$ be the phase of the $k$th agent estimated by taking the Hilbert transform of its position signal, say $x_k\left(t\right)$. The cluster phase of $N$ agents is defined as $q(t) := \dfrac{1}{N} \sum_{k=1}^{N} e^{j\theta_k\left(t\right)}$ , which represents the average phase of the group at time $t$. The term $\phi_k\left(t\right) := \theta_k(t) - q(t)$ is the relative phase between the $k$th agent and the group phase at time $t$.

The level of coordination reached by a human group performing an oscillatory task can be investigated by evaluating the \textit{group synchronisation index} $\rho_g(t)$ introduced in \cite{Alderisio2017, Richardson2012} and defined as follows:
\begin{equation}
\rho_g(t) := \dfrac{1}{N} \left| \sum_{k=1}^{N} e^{j(\phi_k(t) - \bar{\phi}_k)} \right| \quad \in \left[0,1\right],
\end{equation}
where $\bar{\phi}_k$ is $\phi_k(t)$ averaged over time. Closer the synchronisation index is to $1$, higher is the level of synchronisation in the group.

\section{DEEP REINFORCEMENT LEARNING APPROACH}
\subsection{Brief overview}
\label{sec:deep_q_network}
\begin{figure*}[thpb]
	\centering
	\includegraphics[width=0.68\textwidth]{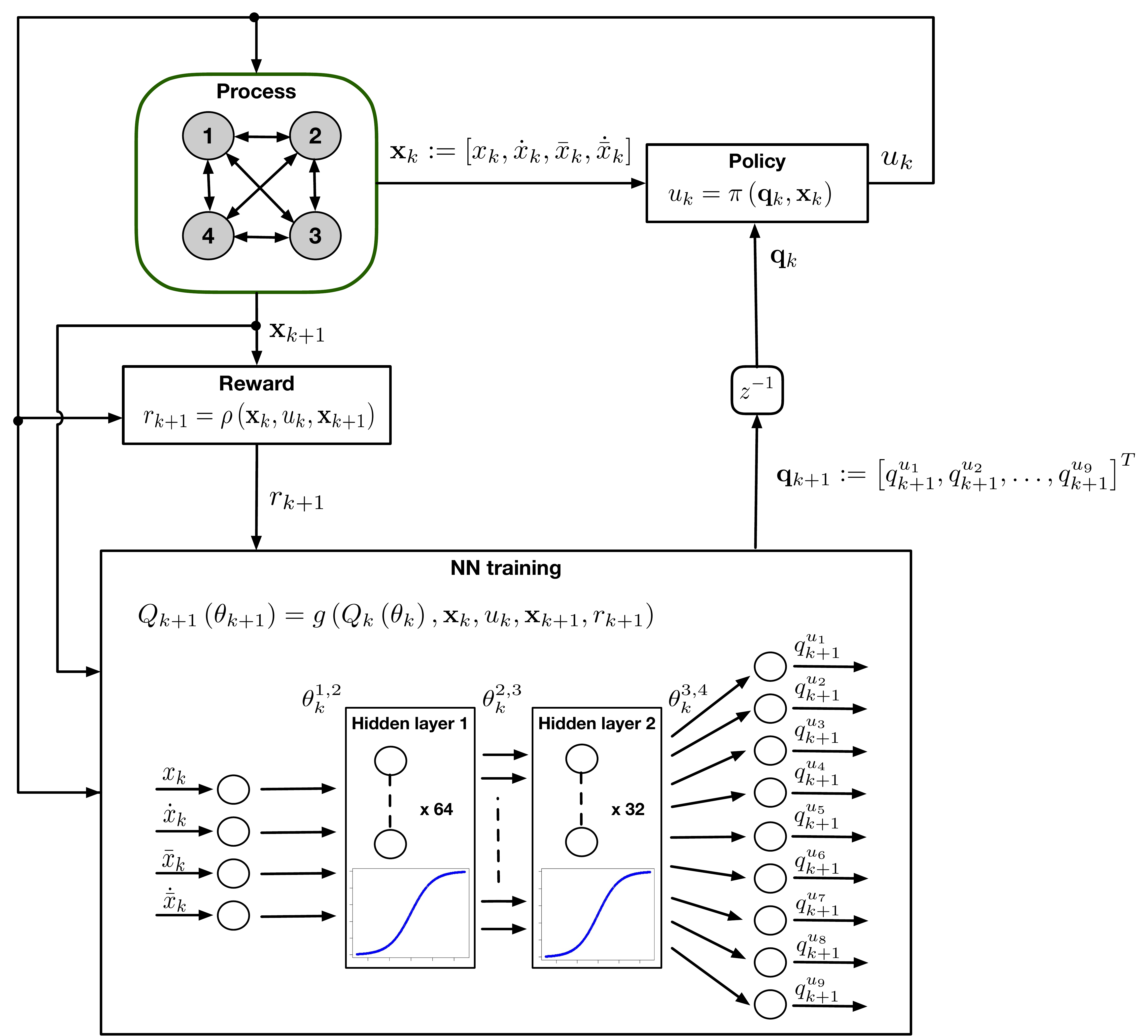}
	\caption{Block scheme of the deep Q network algorithm. At each iteration, the RL controller chooses the control $u$ according to the current neural network and the system's state. The process evolves in a new state generating the reward $r$. The reward, previous and current state are then used as a new sample to train the neural network.}
	\label{fig:dqn_diagram}
\end{figure*}

Reinforcement learning is a machine learning technique in which an agent tries to learn how to behave in an unknown environment taking, in any situation, the best action that it can perform. 
The  problem is formalized by considering a set $X$ of all possible states in which the environment can be (state-space), a set $U$ of all possible actions that the agent can take (action-space) and an auxiliary function, named action-value function (or Q-function), that quantifies the expected return (reward) starting from a specific state and taking a specific action. Through the action-value function, the goal of the learning agent is to iteratively refine its policy in order to maximise the expected reward. Solving a problem with the classical Q-learning approach \cite{Watkins1992} means to iteratively explore all possible combinations between the set $X$ and the set $U$ in order to evaluate them in terms of action-value functions in a tabular form. As this is unfeasible in our group scenario, we use the deep learning control approach shown in Fig. \ref{fig:dqn_diagram} where:
\begin{itemize}
	\item the \textit{state space} is $\mathbf{x} := \left[x, \dot{x}, \bar{x}, \dot{\bar{x}}\right]$ where $\left[x, \dot{x}\right]$ are position and velocity of the CP, while $\left[\bar{x}, \dot{\bar{x}} \right]$ are mean position and mean velocity of the players connected to the target players;
	\item the \textit{action space} is made up of $9$ different values of acceleration in $\left[-4, 4\right]$, empirically chosen looking at the typical human accelerations while performing the same joint tasks;
	\item the \textit{reward function} is selected as $\rho := -\left(x-x_t\right)^2 - 0.1 \left(\dot{x} - \dot{x}_t \right)^2 - \eta u^2$ where $\left[x_t, \dot{x}_t \right]$ are position and velocity of the target player, while the constant parameter $\eta$ tunes the control effort. Maximising a reward function so designed means to minimise the squared error both in position and in velocity between the CP and the target player;
	\item the \textit{policy} $\pi$ according to which the CP chooses the action to take in a specific state is  an $\epsilon$-greedy policy \cite{Sutton2018}. Following that policy, the CP takes the best known action with $\left(1-\epsilon\right)$ probability (exploitation), whereas with $\epsilon$ probability it takes a random action (exploration). The value $\epsilon$ follows a monotonic decreasing function, since as time increases the exploration phase is replaced by the exploitation phase.
\end{itemize}


In particular, we exploit the Deep Q-network (DQN) strategy where an artificial neural network (ANN) is used to approximate the optimal action-value function $Q^*$ defined as:
\begin{multline}
Q^*\left(\mathbf{x}, u  \right) = \max_{\pi} \mathbb{E} \left[ \sum_{m=0}^{\infty} \gamma^m r_{k+m+1} 
| \mathbf{x}_k = \mathbf{x}, u\right],
\label{eq:action-value_function}
\end{multline}
which maximises the expected value of the sum of the rewards $r$ discounted by a positive factor $\gamma < 1$, obtained taking the action $u$ in the state $\mathbf{x}$ following the policy $\pi$ at any time instant $k$.

Training an ANN in order to approximate a desired function (Q-function) means to find the vector of network weights $\mathbf{\theta}$ of the connections between the neurons, iteratively evaluated by back-propagation algorithms in order to minimise a loss function. The loss function is used to measure the error between the actual and the predicted output of the neural network (e.g. mean squared error) (see \cite{Russel2003} for further details).

Contrary to what is done in traditional supervised learning with ANN where the predicted output is well defined, in the Deep Q-network approach the loss function is iteratively changed because the predicted output itself depends on the network parameters $\theta_k$ at every instant $k$. Namely, the loss function is chosen as:
\begin{multline}
L_k\left(\mathbf{\theta}_k\right) = \mathbb{E} \Biggl[ \Biggl(r_k + \gamma \max_{u_{k+1}} Q\left(\mathbf{x}_{k+1}, u_{k+1}, \mathbf{\theta}_{k-1} \right)   - \\
Q\left( \mathbf{x}_k, u_k, \mathbf{\theta}_k \right)\Biggr)^2 \Biggr],
\label{eq:loss_function}
\end{multline}
which represents the mean squared error between the current estimated $Q$ function and the approximate optimal action-value function.

It has been proved that an ANN with a single hidden layer containing a large enough number of sigmoid units can approximate any continuous function, while a second layer is added to improve accuracy \cite{Cybenko1989}. Relying on that, the neural network we considered to approximate the action-value function $Q$ in (\ref{eq:action-value_function}) is designed as a feedforward network with (Fig. \ref{fig:dqn_diagram}):
\begin{itemize}
	\item an \textit{input layer} with $4$ different nodes, one for each state variable $\left[x, \dot{x}, \bar{x}, \dot{\bar{x}} \right]$;
	\item \textit{two hidden layers}, empirically found, made up of $64$ and $32$ nodes each implementing a sigmoidal activation function;
	\item an \textit{output layer} with $9$ different nodes, one for each action available in the set $U$. In the DQN, the network output returns the estimated action-value $q^{u}$ for each possible action $u \in U$ in a single shot reducing in this way the time needed for the training. Then, the action corresponding to the maximum q-value (neuron's output) is chosen as the next control input.
\end{itemize}

Reinforcement learning is known to be unstable or even to diverge when a nonlinear function approximator, such as an ANN, is used to estimate the Q-function \cite{Sutton2018, DeepMind2015}. According to the existing literature, this instability is caused by: (i) the presence of correlation in the sequence of observed states and (ii) the presence of correlation between the current estimated $Q$ and the target network, resulting in the loss of the Markov property. The correlation in the observation sequence is removed by introducing the \textit{experience replay mechanism}, where the observed states used to train the ANN are not taken sequentially but are sampled randomly from a circular buffer. Also, the correlation between the current estimate of the function $Q$ and the target optimal one $Q^*$ is reduced updating the latter at a slower rate.

\subsection{CP Implementation}
According to the reinforcement learning strategy with the Deep Q-network described above, the CP refines its policy according to the system states and the reward received so as to take the best action it can to mimic the target player(s). To implement the DQN as a first step, a feedforward neural network needs to be initialized with random values. The experience replay mechanism is implemented instantiating an empty circular buffer in order to store the system's state at each iteration.

Then at each iteration we have (Fig. \ref{fig:dqn_diagram}):
\begin{enumerate}
	\item the CP observes the process's state $\mathbf{x}_k$ at time instant $k$ and performs an action $u_k$ according to the policy $\pi$, that is an $\epsilon$-greedy policy; 
	\item the process evolves to a new state $\mathbf{x}_{k+1}$ and the CP receives the reward $r_{k+1}$ that measures how good taking the action $u$ in the state $\mathbf{x}_k$ has been;
	\item the new sample $\left\langle\mathbf{x}_k, u_k, \mathbf{x}_{k+1}, r_{k+1}\right\rangle$ is added to the circular buffer and a random batch taken from it is used to train the NN. The training is done through the gradient descend back-propagation algorithm with momentum \cite{Sutton2018} so as to tune the network's weights $\mathbf{\theta}$ in order to minimize the loss function (\ref{eq:loss_function}). We denote the network's weights between the layer $n$ and $n+1$ at instant $k$ as $\mathbf{\theta}_{k}^{n,n+1}$.
\end{enumerate}
The steps above are repeated until convergence is achieved according to the ``termination criterion'':
\begin{equation}
	\left\lVert RMS_{TP,\bar{x}} - RMS_{CP, \bar{x}} \right\rVert \leq \epsilon
\label{eq:convergence}
\end{equation}
where $RMS_{i,\bar{x}}$ is the root mean square error between the position of the player $i \in [TP, CP]$ and the mean position of the group, while $\epsilon$ is a non-negative parameter.

\section{TRAINING AND VALIDATION}

\subsection{Training}
Ideally, data used to train the CP are extracted from real human players playing the mirror game. In our case, due to the lack of a large enough dataset, the data needed to feed the CP during the training are generated synthetically making artificial agents modelling human players perform sessions of the mirror game against each other. We refer to these other artificial agents as Virtual Players (VP) to distinguish them from the CP since they are driven by a completely different architecture which is not based on AI and was presented in \cite{Zhai2017} and improved in \cite{Lombardi2018a}. 
The use of virtual players for training AI based CPs was first proposed in \cite{Lombardi2018b} for dyadic interaction and is applied here for the first time to the multi-player version of the game.

Specifically, the motion of the virtual player used to generate synthetic data is that of a controlled nonlinear HKB oscillator \cite{Haken1985} of the form:
\begin{equation}
	\ddot{x} + \left( \alpha x^2 + \beta \dot{x}^2 - \gamma \right) \dot{x} + \omega^2 x = u,
\end{equation}
where $x, \dot{x}$ and $\ddot{x}$ are  position, velocity and acceleration of the VP end effector, respectively, $\alpha, \beta, \gamma$ are positive empirically tuned damping parameters while $\omega$ is the oscillation frequency. The control input $u$ is chosen following an optimal control strategy aiming at minimising the following cost function \cite{Zhai2016}:
\begin{multline}
	\min_u J\left(t_k\right) = \dfrac{\theta_p}{2} \left( x\left(t_{k+1} \right) - r_p \left( t_{k+1} \right) \right)^2 + \\ 
	\dfrac{\theta_\sigma}{2} \int_{t_k}^{t_{k+1}} \left( \dot{x}\left(\tau\right) - \dot{r}_\sigma \left(\tau\right)\right)^2 \; d\tau + \\
	\dfrac{\theta_v}{2} \int_{t_k}^{t_{k+1}} \left( \dot{x}\left(\tau\right) - \dot{r}_p \left(\tau\right)\right)^2 \; d\tau +
	\dfrac{\eta}{2} \int_{t_k}^{t_{k+1}} u\left(\tau\right)^2  \; d\tau,
\end{multline}
where $r_p, \dot{r}_p$ is the position and the velocity time series of the partner player, $\dot{r}_\sigma$ is the reference signal modelling the desired human motor signature, $\eta$ tunes the control effort, $t_k$ and $t_{k+1}$ represent the current and the next optimization time instant. $\theta_p, \theta_s, \theta_v$ are positive control parameters satisfying the constraint $\theta_p + \theta_s + \theta_v = 1$. By tuning appropriately these parameters, it is possible to change the VP configuration making it act as a leader, follower or joint improviser (more details are in \cite{Zhai2017,Zhai2016}).
In the case of a multi-player scenario, $r_p$ and $\dot{r}_p$ are taken as the mean value of the position and the velocity of the target player's neighbours, that is:
\begin{equation}
	r_p := \dfrac{1}{M}\sum_{j=1}^{M} x_j; \quad \dot{r}_p := \dfrac{1}{M}\sum_{j=1}^{M} \dot{x}_j,
\end{equation}
where $M$ is the number of neighbours and $x_j$ and $\dot{x}_j$ are the position and the velocity of the $j$th neighbour.

The reference signal $\dot{r}_\sigma$ captures in some way the desired human kinematic features that the VP has to exhibit during the game. In \cite{Lombardi2018a} we developed a methodology based on the theory of stochastic processes and observational learning to generate human-like trajectories in real time. In particular, a Markov Chain (MC) was derived to capture the peculiar internal description model of the motion of a human player simply observing him/her playing sessions of the mirror game in isolation.

\begin{figure}[thpb]
	\centering
	\includegraphics[width=0.9\columnwidth]{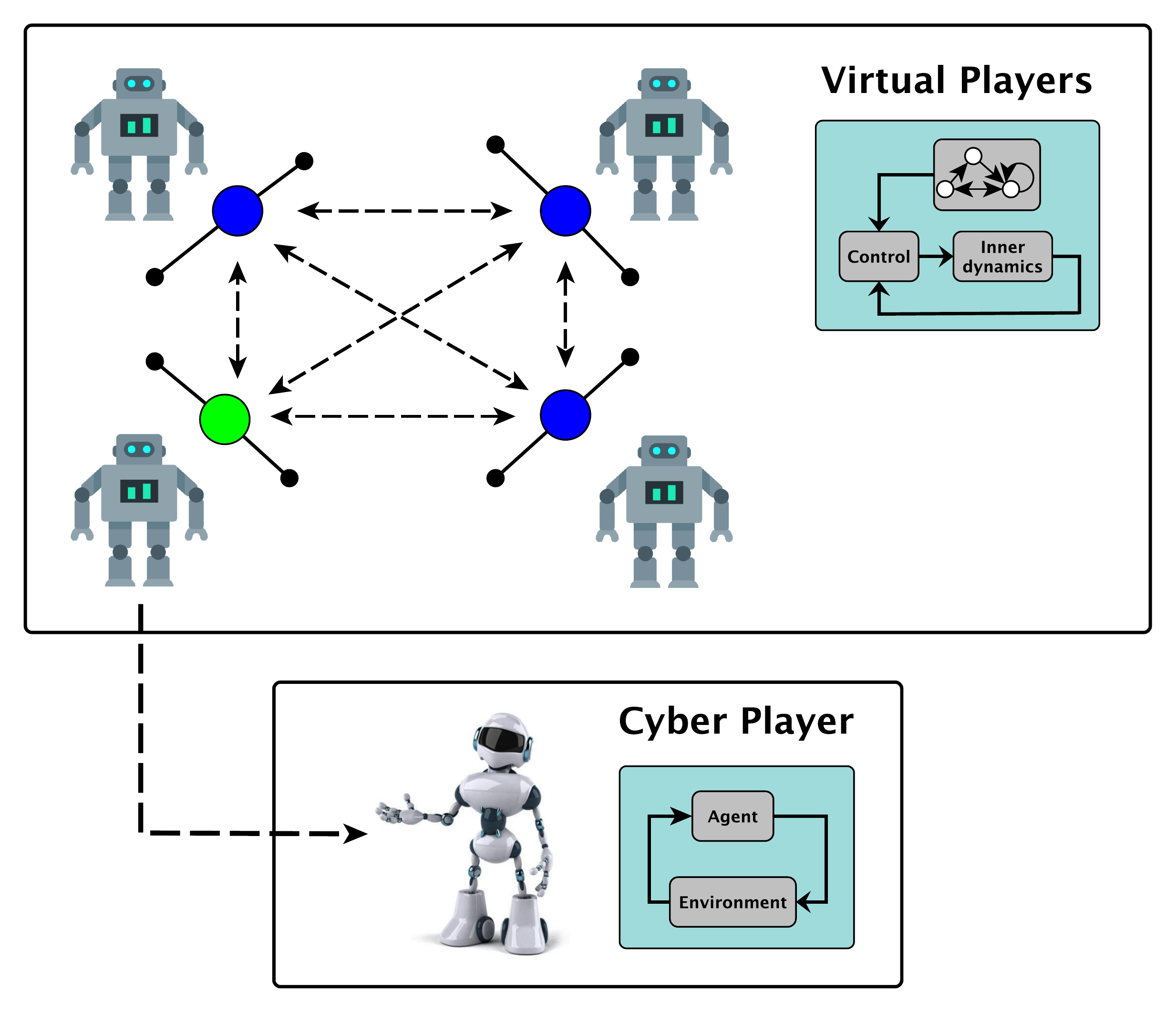}
	\caption{Architecture used to train the CP to mimic the target player (ball in green). The group is simulated by $4$ VPs playing the mirror game in joint improvisation and driven by the cognitive architecture based on optimal control and Markov chains. The CP, driven by a deep reinforcement learning approach, receives in real time the data from the group and learns how to interact in order to replace the player target.}
	\label{fig:schema_group}
\end{figure}

To train the CP to coordinate its movements in the group like the virtual player target does, a group of $4$ different virtual players interconnected in a complete graph were used (Fig. \ref{fig:schema_group}). In particular we selected four Markov models (one for each player) of different human players which were parametrized in \cite{Lombardi2018a}. Without loss of generality, VP $4$ was taken as the target player the Deep Learning driven CP has to mimic.

The parameters proposed for the control architecture of the VPs were tuned experimentally as follows: $\alpha=1, \beta=2, \gamma=-1,\omega=1$ for the inner dynamics, $\theta_p = 0.8, \theta_\sigma = 0.15, \theta_v = 0.05, \eta=10^{-4}$ and $T=0.03s$ for the control law.
The experience replay in the CP algorithm was implemented with a buffer of $200,000$ elements, batches of $32$ sampled states were used to train the feedforward neural network at each iteration. A target network updated every $150$ time steps was considered in the Q-function, with a discount factor $\gamma = 0.95$.

\begin{figure}[thpb]
    \centering
	\includegraphics[width=0.99\columnwidth]{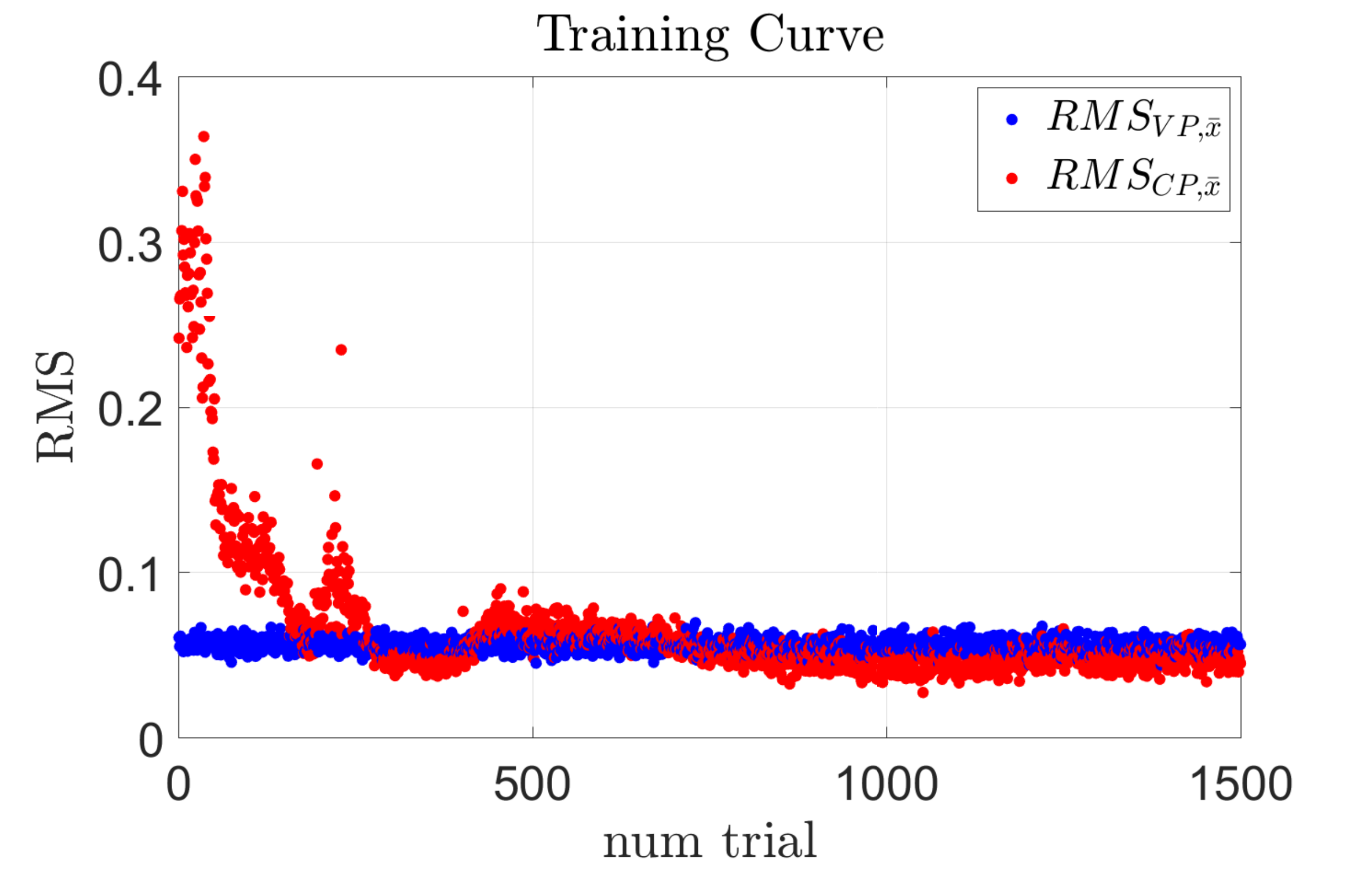}
	\caption{Training curve showing the convergence of the algorithm. The root mean square error in position (y-axis) is reported for each trial (x-axis) both for the VP (in blue) and the CP (in red). $\bar{x}$ represents the mean position of the target player's neighbours.}
	\label{fig:training_curve}
\end{figure}

The training stage was carried out on a Desktop computer having an Intel Core i7-6700 CPU, 16 GB of RAM and 64-bit Windows operative system. It took $1,500$ trials  of $500$ observations each to converge (around $8.5$ hours) according to the criterion (\ref{eq:convergence}). In Fig. \ref{fig:training_curve} the training curve is reported showing for each trial the RMS between the VP and the group (in blue), and between the CP and the group (in red). Convergence is reached in about $1,000$ trials on.

\subsection{Validation}
\begin{figure*}[thpb]
	\centering
	\includegraphics[width=\textwidth]{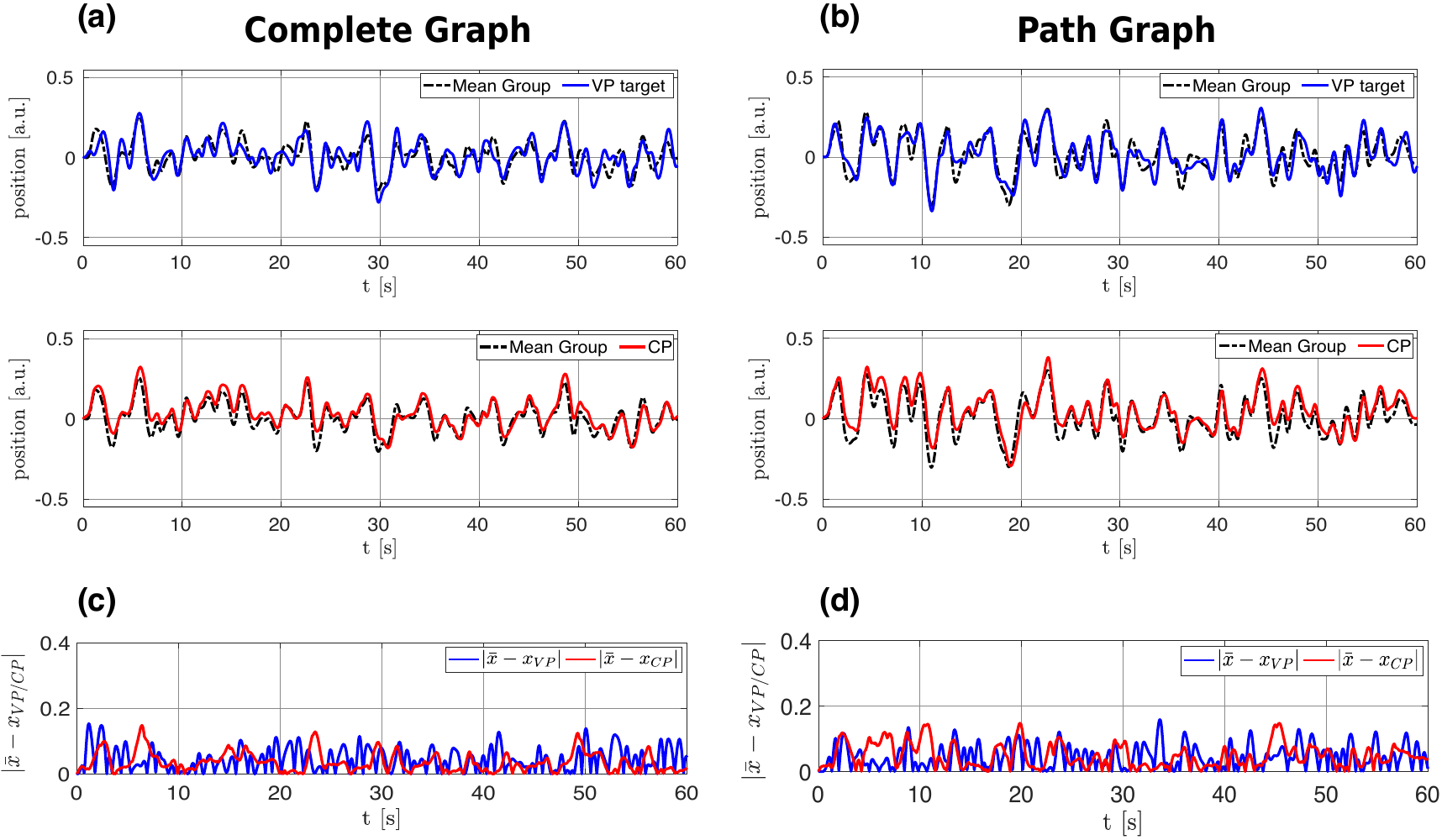}
	\caption{Position time series of group sessions. \textbf{(a)}-\textbf{(b)} The position trajectories of the VP target (in blue) and of the CP (in red) are reported together with the mean position of the neighbours (in dashed back). \textbf{(c)}-\textbf{(d)} Relative position error evaluated between the mean position and the position of the VP target (in blue) and the CP respectively. 
	Two different topologies are depicted: \textbf{(a)}-\textbf{(c)} the complete graph and \textbf{(b)}-\textbf{(d)} the path graph.}
	\label{fig:validation_CP_groups}
\end{figure*}

To show that the CP is effectively able to emulate the VP target when engaged in a group, training was carried out by considering a group described by a complete graph, we then validated the performance of the CP when interacting over different topologies. Specifically we used the ring graph, path graph, star graph (described in Sec. \ref{sec:background}) with node $3$ as center. For the sake of brevity, in Fig. \ref{fig:validation_CP_groups} only the validation for the complete and the path graph are reported. The performance of the CP has been evaluated by comparing its behaviour with that of the target virtual player it was trained to mimic. The CP (in red) successfully tracks the mean position of the group (dashed line in black) being able to mimic the target player it has been trained to imitate (in blue) [panel (a)]. The relative position error (RPE) defined as
\begin{equation}
    RPE =
    \begin{cases}
        \!\begin{aligned}
         & \left(\bar{x}(t) - x_{VP/CP}(t) \right)sgn\left(\dot{\bar{x}}(t)\right) \\
         & \quad \text{if $sgn\left(\dot{\bar{x}}(t)\right)=sgn\left(x_{VP/CP}(t)\right)\neq0$}
         \end{aligned} \\
         \left|\bar{x}(t)-x_{VP/CP}(t)\right|, \qquad \text{otherwise}
    \end{cases}
\end{equation}
has been also evaluated between the VP target and the mean position of the neighbours and compared with the relative position error between the CP and the same mean position [panel (c)]. Both the errors are very small and with comparable mean values.

Similar considerations can be done for behaviour of the CP when the group topology is a path graph as shown in panels (b) and (d).

The key features of the motion of the CP and the VP it has been trained upon are captured by the following metrics: 1) relative phase error defined as $\Delta \Phi = \Phi_{\bar{x}} - \Phi_{CP/VP}$, 2) the RMS error between the position of the CP (or VP) and that of the group mean position, and 3) the time lag which describes the amount of  time shift that achieves the maximum cross-covariance between the two position time series. This can be interpreted as the average reaction time of the player in the mirror game \cite{Alderisio2017}. 

The metrics described above were evaluated performing $20$ trials and reporting both the mean value and the standard deviation for both the complete graph (Tab. \ref{tab:metrics_complete}) and the path graph (Tab. \ref{tab:metrics_path}). It is possible to notice that all indexes show a remarkable degree of similarity between the motion of the CP and that of the target VP.

\begin{table}[thpb]
\caption{Metrics are reported for $20$ trials of the multiplayer mirror game both in complete graph (a) and in path graph (b). The relative phase, the RMS and the time lag are evaluated between the mean of the neighbours with the CP (first column) and with the VP target (second column).}
\begin{subtable}{\columnwidth}
\centering
\begin{tabular}{|c||c|c|}
\hline
\textbf{Metric} & \textbf{CP} & \textbf{VP target} \\
\hline
Relative phase & $-0.1680 \pm 0.0513$ & $-0.0498 \pm 0.0948$ \\
\hline
RMS & $0.0443 \pm 0.0054$ & $0.0556 \pm 0.0037$ \\
\hline
Time lag & $-0.1140 \pm 0.0157$ & $-0.0165 \pm 0.0357$ \\
\hline
\end{tabular}
\caption{Complete Graph}
\label{tab:metrics_complete}
\end{subtable}
\vspace{10pt}
\begin{subtable}{\columnwidth}
\centering
\begin{tabular}{|c||c|c|}
\hline
\textbf{Metric} & \textbf{CP} & \textbf{VP target} \\
\hline
Relative phase & $-0.2296 \pm 0.0788$ & $-0.1589 \pm 0.0982$ \\
\hline
RMS & $0.0555 \pm 0.0047$ & $0.0597 \pm 0.0035$ \\
\hline
Time lag & $-0.1035 \pm 0.0153$ & $-0.0675 \pm 0.0236$ \\
\hline
\end{tabular}
\caption{Path Graph}
\label{tab:metrics_path}
\end{subtable}

\end{table}

\begin{figure}[thpb]
	\centering
	\includegraphics[width=0.9\columnwidth]{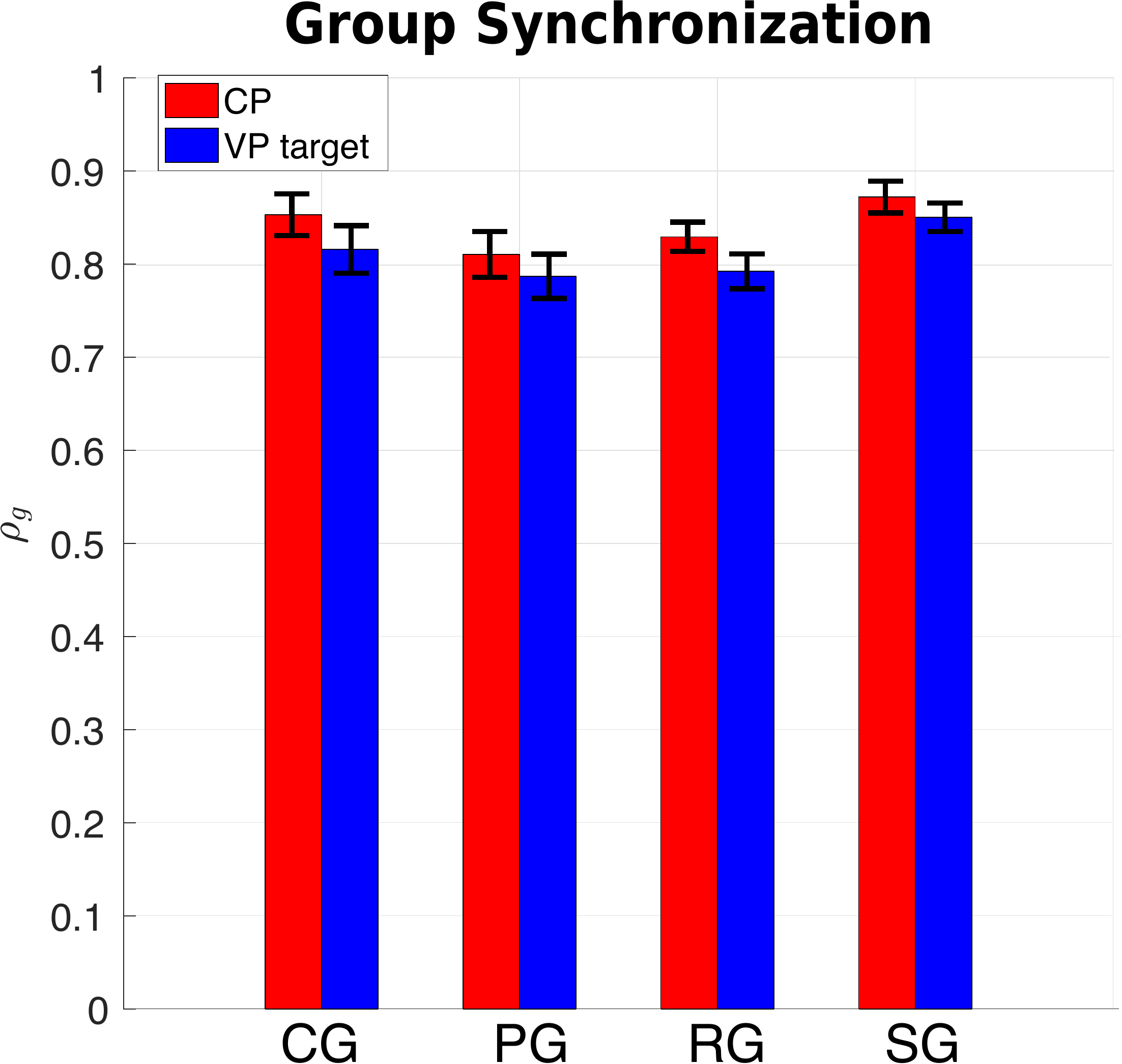}
	\caption{Histogram reporting the group synchronization level reached by the group both with the VP target (blue bars) and with the CP (red bars). Different topologies were implemented during the validation: complete graph (CG), path graph (PG), ring graph (RG) and star graph (SG) with player 3 as center node.}
	\label{fig:group_sync}
\end{figure}

For further evidence, in Fig. \ref{fig:group_sync} the group level synchronization is reported for each tested topology. Despite the different topologies, the presence of the CP does not alter the group dynamics when the CP is substituted to the VP it was trained upon. We notice that the level of coordination varies with the topology, confirming what found in \cite{Alderisio2017}. Specifically, in \cite{Alderisio2017} as confirmed by Fig. \ref{fig:group_sync}, the complete and the star graph were found to be associated with the highest level of synchronization.

\addtolength{\textheight}{-8cm}   

\section{CONCLUSIONS}
\label{sec:conclusions}
In this work we addressed the problem of synthesizing an autonomous artificial agent able to coordinate its movements and perform a joint motor task in a group setting. In particular, a multiplayer version of the mirror game was used as a paradigmatic task where different individuals have to synchronize their  oscillatory motion. To achieve our goal and overcome the limitations of previous approaches, we introduced a deep reinforcement learning control algorithm in which a feedforward neural network is used to approximate the nonlinear action-value function. The DQN allowed us to overcome the limitations of the Q-learning approach presented in \cite{Lombardi2018b} which is impractical when the state space becomes too large, as in the case of multiplayer coordination tasks. The effectiveness of the cyber player trained upon a target group member was shown by comparing its performance when playing in groups with different interconnection topologies. The numerical validations show the effectiveness of our approach. Ongoing work is being carried out to validate the behaviour of the CP when interacting with a real group of people through the experimental platform Chronos we presented in \cite{Alderisio2017a}.



\end{document}